\documentclass[11pt]{article}
\begin{document}

\setlength{\textheight}{21truecm}
\setlength{\textwidth}{16.0truecm}
\setlength{\oddsidemargin}{0.5truecm}
\setlength{\evensidemargin}{0.5truecm}
\setlength{\topmargin}{0.0truecm}

\renewcommand{\baselinestretch}{1.0}
\newcommand{\ba}{\begin{eqnarray}}
\newcommand{\ea}{\end{eqnarray}}
\def \colthree {\mbox{$\bar{\mathbf{3}}_c$}}
\def \colsix {\mbox{$\mathbf{6}_c$}}
\def\rmb#1{{\bf #1}}
\def\lpmb#1{\mbox{\boldmath $#1$}}
\def \CScs {\mbox{${\mathbf{C}_{S}$}}}
\def \CSthreeone {\mbox{$\bar{\mathbf{3}_1$}}}
\def \CSsixone {\mbox{$\mathbf{6}_1$}}
\def \CSthreezero {\mbox{$\bar{\mathbf{3}_0$}}}
\def \CSsixzero {\mbox{$\mathbf{6}_0$}}
\def \CSthreebarone {\mbox{$\bar{\mathbf{3}}_{1/2}$}}
\newcommand{\SpSp}[2]{ \mbox{$\vec{\sigma }_{#1}.\vec{\sigma }_{#2}$}}
\newcommand{\lala}[2]{ \mbox{${\vec{\lambda}_{#1}\cdot
      \vec{\lambda}_{#2}}$}}
\newcommand{\lalaq}[1]{ \mbox{$\frac{\vec{\lambda}_{#1}}{2} \cdot
      \frac{- \vec{\lambda}_{\bar{q}}^*}{2}$}}

\begin{center}
{\Large \textbf{On Quarks and Flavour Symmetry}}\\
\vspace{25mm}
{\large H. H\o gaasen}\\
{\em Department of Physics
University of Oslo\\
Box 1048  NO-0316 Oslo Norway\\
hallstein.hogasen@fys.uio.no}\\

\vspace{10mm}

and\\

\vspace{10mm}

{\large P. Sorba} \\
{\em LAPTH \footnote{Laboratoire d'Annecy-le-Vieux de Physique
Th\'{e}orique, UMR 5108}, Universit\'e de Savoie, CNRS
\\
9 chemin de Bellevue, B.P. 110, F-74941 Annecy-le-Vieux Cedex,
FRANCE\\
sorba@lapp.in2p3.fr}


\end{center}

\vspace{20mm}

\begin{abstract}
Hadronic spectroscopy can be introduced to students and developed
rather far
without requiring SU(N) flavour symmetry. In such a "minimalist"
presentation, we
are naturally led to comment and clarify the concept of the
"generalized" Pauli principle.
\end{abstract}

\vfill \vfill

\clearpage
\pagestyle{plain}
\baselineskip=18pt

\section{Introduction}
Internal symmetry groups have a glorious place in the history of
physics.
One of the highlights was the discovery by Fermi that low energy
pion-nucleon scattering
  is dominated by a single resonance of spin 3/2.
By treating the nucleons as an isospin doublet , the pions as an
isospin
triplet
and the resonance $\Delta$ as a state with isospin 3/2, Fermi and his
collaborators proved the dynamical fact that, to a good
precision, isospin is
conserved in strong interactions.\\
Another highlight was Gell-Mann's prediction of the spin 3/2 $\Omega^-$
at the Geneva conference in 1962. The subsequent discovery of this
state
at Brookhaven in 1964 made everyone confident that the $\it {flavour-
group}$
SU(3)
was as relevant for strong interactions as the isospin group.\\
There is however a difference between the two cases we have mentioned:
Fermi's discovery was a discovery in the dynamics of particles while
Gell-Mann's prediction is usually presented as coming from the
assumption
  that baryons should fall
into specific representations of SU(3)-flavour, namely the 8
dimensional one
for the lightest spin 1/2 baryons, the 10 dimensional one
for the lightest spin 3/2 baryons.\\
Experimental results from reactions involving the assignment of
mesons
and
baryons
to irreducible representations of SU(3) came later, and showed that
the
concept of
  "broken flavour symmetry" was useful.\\
Subsequently the flavour symmetry groups SU(4) and SU(5) have been
invoked
to classify the multiplicities of states when charm and bottom quantum
numbers are added.\\
But let us stress that the assignment of a multiplet of particles
to a group representation is in itself an empty statement. Its
usefulness depends on the group to be an (almost) symmetry group.
When Fermi's team discovered that the isospin group -
historically, this group was first supposed to be R(3) before
becoming SU(2) - it was by studying reactions. If a pion-nucleon
state is a linear combination of isospin I = 1/2 and I = 3/2, only
two independent amplitudes $A_{1/2}$ and $A_{3/2}$ will describe
all $\pi N \rightarrow  \pi N $ reactions \footnote{As $A_{3/2}$
dominates at low energy, the analysis
of data were quite simple.}: this turned out to be correct.\\
Experiments must decide to which extent flavour symmetry is a useful
concept.
 From the multiplication table of SU(3):
\begin{center}
  $8 \otimes 8 = 1 \oplus 8 \oplus 8 \oplus 10 \oplus \overline{10}
\oplus
27$\\
\end{center}
  it follows that the 64 reactions we have in "quasi elastic"
$2\leftrightarrow 2$
   reactions between octet states are given by only six independent
amplitudes.
    From the same
  table we see, as the representation 10 occurs only once, that in the coupling of a decuplet to two octets there
is
only one
  amplitude. This is similar to the decay of $\Delta \rightarrow N
  \pi$ in the isospin symmetric - SU(2) - case. So, if one has baryons made
  from u, d and s quarks, the
  assumption
of flavour symmetry
   relates the coupling constant in $\Delta ^{++} \rightarrow P \pi
^{+}
$ to
   the coupling constants in
   $\Sigma ^{*0} \rightarrow \Lambda \pi ^0 $, $\Sigma ^{*0}
\rightarrow
\Sigma ^{+}
    \pi ^{-} $
   and $\Xi ^{*0} \rightarrow \Xi ^{0} \pi ^0 $. Experimentally they
come out
    correctly within $10\% $ .\\

1964 was the year Gell-Mann and Zweig invented quarks.\\
Right from the start Zweig was a fervent believer in the
objective
existence of quarks
(that he called aces) as the fundamental constituents of hadrons.
With three flavours of  aces (quarks) mesons came in $q\overline{q}$
nonets.
There was no need to invoke a "magic mixing" between a flavour octet
and a
singlet; to explain that one of the vector mesons decayed into
$K\overline{K}$ pairs was simply the reflection of the fact that it was
composed of
$s\overline{s}$.
If some "magic" is there , it is rather in the structure of the spin
zero
mesons.\\

In the years before quarks were generally accepted as physical beings
in
the physics community, flavour symmetry arguments were ( and often
still are! )
  regarded as more high-brow than arguments simply based on quarks.
But today hardly anybody denies that quarks exist.
It is easy to criticize the (nonrelativistic) quark model where
baryons
are made from three valence quarks and mesons from a quark-antiquark
pair, but
it cannot be denied that it gives a set of rules to classify
hadron
states, and also to
calculate many of their static properties which are quite illuminating.\\

However, as we will see in the following, we do not need
any flavour SU(N) group either to classify hadrons or to estimate
many of their physical properties. Hereafter, the fundamental
symmetry groups we will use are the group of rotations in the three
dimensional space represented by the SU(2) spin  group ( together
with  the O(3) orbital angular momentum one in the case of  excited
states), and  the
colour SU(3) group.\\
Indeed, the SU(3) symmetry group of fundamental importance is
not the group acting over flavours, but
the group acting over colour space. It is a group that is gauged and
thereby introduces gluonfields while defining QCD.
The (yet unproven) dogma that free particles are colourless, i.e.
transform
as a singlet under the colour SU(3) group,  implies that the
most economical configurations are those involving three quarks
(baryon) or a quark-antiquark pair (meson).  This comes from the
decomposition of (colour)
SU(3) representations:
\begin{center}$ 3 \otimes 3\otimes 3 = 1 \oplus 8 \oplus 8 \oplus 10 $
    \hspace{10mm}$
and $\hspace{10mm}   $ 3 \otimes 3 = 6 \oplus \overline{3} $
\hspace{10mm}
$3\otimes \overline{3} = 1 \oplus 8 $\end{center}
where the representations 6 and $\overline{3}$  respectively
involve symmetric
and antisymmetric combinations of the two colour indices.\\
The second essential ingredient we need is the Pauli
principle.  Of course it is of importance only for baryons
\footnote{We do not consider mixing with multiquark  states. For
mesons we need a generalization of the Pauli principle which follows from local
field theory through the CPT operation; but this point will not be
discussed hereafter where we limit to (usual) baryons.}
, and it is actually
this class of hadrons which deserves to be considered for our
purpose.
Indeed, at least with regard to their classification,
$q\overline{q}$ mesons
with a given spin and parity are simply gathered in multiplets of
dimension $N^2$, $N$ being the number of different quark flavours. \\
Note that such  three quark states are colour-antisymmetric  under
the exchange of two quarks. Then, the Pauli principle tells us that
a state involving two identical
quarks, i.e. quarks with identical flavours, must be symmetric in
the other quantum numbers.\\

We will start by showing in Sections 2 and 3 that
the multiplets of low-lying as well as excited baryonic resonances
can be directly constructed  with the above hypotheses and tools
at hand. In Section 4,
expressions for S-wave three quark states only satisfying
permutational
symmetry imposed by the Pauli principle are given and used  to
compute baryon mass splittings through (colour) magnetic interactions between
the constituent quarks; their magnetic moments are also  obtained.

However the basis constructed in Section 2
 does not appear suitable for computations of physical
quantities involving flavour changing forces, such as, for instance,
semi-leptonic
decays. A convenient way to circumvent this difficulty
is to impose complete symmetry under permutations of the three
constituents within the baryon wave function rather than the - partial-
symmetry
dictated by the Pauli principle: this prescription appeals to
the so-called "generalized" Pauli principle, which, as is well known,
does not introduce any extra physical assumption. The corresponding
developments are presented in Section 5.

We conclude by
commenting on the direct connection between irreducible
representations of the unitary groups and their properties under the
permutation group, which explains, in our opinion, their perfect
adaptability to the classification of baryons.

In order not of overload the text and keep clear the main idea of
this note, we have decided to treat in  appendices the following
points. Appendix A is devoted to a general and explicit
construction of the (three state particle) Jacobi coordinates. In
our knowledge, such coordinates have already been explicated in
the case of the harmonic oscillator  with only one coupling
constant, but not with a different coupling associated to each
couple of constituents.

Then we give a proof, in Appendix B, of the equivalence between the
two Hilbert space descriptions introduced in Section 4 and 5
respectively, which involve
states satisfying  partial permutation symmetry
 (simple Pauli principle) for the former, and  total permutation
symmetry
(generalized Pauli principle) for the later.  The Fock space emerges
rather naturally from this construction, and we  show in Appendix C
how it can be used to represent baryon states. Finally, we give in
Appendix D the explicit wave-functions of the eight $S=1/2$ low-lying
baryons in the three bases successively introduced: the permutational
"partial"
symmetric basis (cf. Section 4), the totally symmetric basis ( cf.
Section 5) and the Fock space basis, this last one combining the
advantages of the  other two representations: simplicity of the
expressions and complete symmetry.\\
As this paper can be partly regarded as a set of lecture notes, there is
an almost empty reference list. We apologize for this but we realize that to do
 justice to all the people who developed
 the subject, we would need a list longer than the article.

\section{The ground state of three quarks}

This is the subject of most introductory courses on elementary
particles.
So, we will treat this first. Although the following is known, it is
apparently not well known, as we realized by looking up at lectures
notes on the web. Among many popular monographs, we found only one \cite{MartinShaw}
which shares the approach presented below.\\
We now forget about colour, remembering that each pair of quarks are
antisymmetric in
colour so that the baryon states must be symmetric in the other
degrees
of
freedom.
This was for a long time called the Dalitz symmetric quark model.
Note at this point that it was realized \cite{Franklin:68} that if quarks were fermions
such that there existed an hidden quantum number - which later turned
out to be the colour - justifying the symmetric quark model, then the
multiplicity of the ground states would follow independently of any
flavour symmetry group.\\
In the ground state where no angular momentum is involved, the only
degree
freedom of each quark is the spin. As a pair of identical quarks must
be
symmetric under interchange  it must be symmetric in spin: the spin
of
the pair is one.\\
Integrating out the spatial degrees of freedom, we are left with a
  Hamiltonian over the flavour-spin space of the quarks.
   \footnote{The following arguments are not restricted to
nonrelativistic
quantum mechanics.}.
Now let us  count the ground states:\\
Suppose that we have N flavours (u, d, s, c,..) and we choose three
of these to make  baryons.\\
First for three identical flavours:\\
There are N ways of choosing these baryons made up of three identical
quarks.
Each pair has S=1 so all have total spin 3/2.\\
Two identical flavours:\\
There are N(N-1) ways of choosing two identical and one different
from the two first chosen.\\
The identical quarks are coupled to spin one, and the total spin is
S=3/2 or S=1/2.\\
All three quarks different:\\
There are N(N-1)(N-2)/6 ways of choosing three different
flavours.The Pauli principle imposes no restriction here, so the
total spin can be 3/2 or 1/2, the multiplicity of S=1/2 is two.\\
So the number N(3/2) of flavour states with total spin 3/2 is:
\begin{center}
N(3/2) =N+N(N-1)+ N(N-1)(N-2)/6 = N(N+1)(N+2)/6\end{center}
and the number of flavour states with spin 1/2  is:
\begin{center}
N(1/2)= N(N-1)+2N(N-1)(N-2)/6 =N(N+1)(N-1)/3\end{center}
We immediately see that we have found the same multiplicities as is
commonly  inferred from the dimensions of the representations of the flavour
symmetry group
SU(N).\\
For N=3 we have an "eightfold way", more precisely an octet of S=1/2
states and a decuplet of S=3/2 states. \\
But we also realize that these multiplicities have nothing to do with
the existence of an internal
symmetry group, they have their sole origin in the
Pauli principle. Not only u,d,s, quarks, but also u,c,b, or any
triplet of flavour quarks will provide with an octet and a decuplet.\\
Now, let us count the total number of such quantum mechanical states.
Since there are 2S+1 states in a spin S representation, one gets:

 \begin{center} N(total)= 4 N(N+1)(N+2)/6 +2 N(N+1)(N-1)/3 = 2
     N(N+1)(2N+1)/3\end{center}
and that is exactly the dimension of the SU(2N) completely symmetric
representation arising from the tensorial product of three 2N
dimensional fundamental SU(2N) representations. Taking as an example N=
3, we indeed get 56 states, that is the dimension of the
corresponding symmetric representation of the usually called
"flavour-spin" SU(6) group: we will comment later on this point,
in direct connection with the "generalized" Pauli principle.

\section{Excited states}

Let us now turn our attention to the construction of P-wave baryon
states, to which we will restrict our presentation in order not to
overload this note. These negative parity baryons are associated to
an
 L=1 orbital momentum. This unit of O(3) orbital angular momentum
 stands usually in one of the two quark relative coordinates, and the
 most common model which is well adapted to represent such an effect
 is that of the harmonic oscillator. The reason for treating confining
 forces between quarks using harmonic oscillator potentials is simply
 that the center of mass motion can be separated out using Jacobi
coordinates.
 Note that this choice  has no importance for our purpose,
 namely the $\it number$ of exited states.\\
 We will use as a Hamiltonian
\begin{equation}
H_{\rm h.o.}=\sum_i( m_i + p_{i}^2 / 2m_i)
+ \sum_{i<j}  (k/2) \cdot (\overrightarrow r_i -\overrightarrow
r_j)^2
)
\label{IKHam}
\end{equation}
 where $\overrightarrow{r_{{i}}}$, i=1,2,3, denote the respective
positions of
 the three quarks. Although we do not start with  a more general
Hamiltonian
 $H_{ gen}$, we can obtain
the general solution for the spectrum by first using $H_{\rm h.o.}$
to get states of a harmonic oscillator basis
and then perturb them with the perturbation $H_{\rm gen}$-$H_{\rm
h.o.}$.
The energy of the levels will change with respect to the ones of the
harmonic
oscillator,
but their number will be the same.\\
In terms of the Jacobi coordinates $\overrightarrow{\rho },
\overrightarrow{\lambda }$ and the center of mass coordinates
  $\overrightarrow{R_{cm}}$,
the Hamiltonian separates into $H_{\rm h.o.}= \sum_i m_i  +H_{cm}+
H_{\rm \rho}+ H_{\rm \lambda}$ with
\begin{eqnarray}\label{sing:eq:ham}
&H_{\rm cm}=P^2 / 2M,\quad
M=\sum_i m_i,\quad &\overrightarrow P =\sum_i \overrightarrow
p_i,\quad
\nonumber\\
&H_{\rm \rho}=p_{\rho}^2 / 2m_{+}
+  3 k/2 \cdot (\overrightarrow \rho )^2 \\
&H_{\rm \lambda}=p_{\lambda}^2 / 2m_{\lambda}
+  3 k/2 \cdot (\overrightarrow \lambda )^2  \nonumber.
\end{eqnarray}
The explicit expressions of $m_{+}$ and $m_{-}$ are given in
Appendix A, where a detailed construction of Jacobi coordinates is
developed, and also generalized to the  case where we have a
different coupling $k_{ij}$ for each quark pair $q_{i}q_{j}$ in the
potential term.\\
For notational simplicity, we consider here the equal mass case, i.e.
$m_{i}=m$
to which correspond: $m_{+}=m_{-}=m$ and
   the relative coordinates $\overrightarrow{\rho }$ and
   $\overrightarrow{\lambda }$ reduce to :

 $$\overrightarrow{\rho }= (\overrightarrow{r_{{1}}} -
    \overrightarrow{r_{{2}}}) /\sqrt{2}$$
 $$\overrightarrow{\lambda }= (\overrightarrow{r_{{1}}} +
    \overrightarrow{r_{{2}}} -2 \overrightarrow{r_{{3}}}) /\sqrt{6}$$
So, let us separately study the different configurations:

 i) $\textbf{Baryons  made of three flavour identical quarks
 $qqq$}$:

    Consider the oscillator relative to  $\overrightarrow{\rho }$: it
    is antisymmetric in the exchange (1) -(2). The doublet made with
    the first two quarks $qq$ , that is of quarks in position (1) and
    (2), has automatically  spin $S= 0$, in order  for
the total spin and orbital
    momentum part to be symmetric ( Pauli principle ). It follows that in this
$\overrightarrow{\rho }$
    configuration, the total spin of the baryon is $S = 1/2$, and the
    spin/orbit part of the wave function reads,for $Sz= +1/2$ and up to
    a normalization factor:

     ($\overrightarrow{r_{{1}}}$ -
    $\overrightarrow{r_{{2}}}$) ($ \uparrow \downarrow \uparrow $ -
    $ \downarrow \uparrow \uparrow $ ) + sym.

    Replacing now $\overrightarrow{\rho }$ by $\overrightarrow{\lambda
    }$, which is symmetric in the first two quarks, one easily deduces
    that the corresponding $qq$ doublet has a (symmetric) spin $S= 1$.
However
    in the product $S= 1$ by $S=1/2$, the totally spin symmetric $S=
    3/2$
part
cannot
    provide, when combined with the (not completely symmetric)
    $\overrightarrow{\lambda }$, a completely symmetric wave function.
    It follows that the only possibility  for the resulting baryon
is to have spin 1/2.
    Moreover, one notes that the spin/orbit part of the wave function
    reads:

    ($\overrightarrow{r_{{1}}}$ +
    $\overrightarrow{r_{{2}}}$ -2 $\overrightarrow{r_{{3}}}$) ($ 2
    \uparrow \uparrow \downarrow $ - $ \downarrow \uparrow \uparrow $
    -  $ \uparrow \downarrow \uparrow $) + sym.

    which is exactly the  same, up to a scale factor, as the one obtained
    just above for the  $\overrightarrow{\rho }$ case.

    Therefore, for $qqq$ configurations, the only solution is given by
    the $\overrightarrow{\rho }$ oscillator with two quark spin one
    $\it{or}$
    by the  $\overrightarrow{\lambda }$ oscillator with two quark
    spin zero,
    and total spin $S = 1/2$.\\

    ii) $\textbf{Baryons made of two (and only two) identical flavour quarks
$qqq'$}$:

    Now the Pauli principle imposes permutation symmetry only in the
    first two quarks $qq$. Note that, as explicated in Appendix A,
    the antisymmetry ( resp. symmetry ) of $\overrightarrow{\rho }$ ( resp.
  $\overrightarrow{\lambda }$ ) persists when $m_{1}=m_{2}=m$  and
  $m_{3}=m'$ with $m$ and $m'$ different. Then one gets:

    -  with the $\overrightarrow{\rho }$ oscillator: the $qq$ spin part
must
be zero, leading for the (qqq') baryon to $S=1/2$, with in the wave
    function a $qq$ spin/orbit part proportional to:

    ($\overrightarrow{r_{{1}}}$ -
    $\overrightarrow{r_{{2}}}$) ($ \uparrow \downarrow $ -
    $ \downarrow \uparrow $ ) + sym.

    - with the $\overrightarrow{\lambda }$ oscillator: the $qq$ spin part
    is now one, allowing the total spin to be $S = 3/2$ and $S = 1/2$.\\

    iii)$\textbf{ Baryons made of three different flavour quarks
    $qq'q"$}$:

    Now there is no restriction imposed by the Pauli principle.
    It follows that  with the $\overrightarrow{\rho }$ as
    well as with the  $\overrightarrow{\lambda }$ oscillator, one can
    get one
    baryon with total spin $S = 3/2 $ and two baryons of total spin $
    S = 1/2$ for each triplet of flavours.\\

    As a conclusion, let us count the number of states  of three
    different flavours:

    - with $S = 3/2$:
    there are 6 states of $qqq'$-type with the $\overrightarrow{\lambda
    }$ oscillator, one state $qq'q"$ with each of the
$\overrightarrow{\rho }$ and
     $\overrightarrow{\lambda }$ oscillator, that is a total of 8
states.

     - with $S = 1/2$:
     there are 3 states of $qqq$-type made from the
$\overrightarrow{\rho
     }$ osc. (and partial $qq$ spin $S = 0$) $\emph{or}$ from the
$\overrightarrow{\lambda }$ osc.
     ( and partial $qq$ spin $S = 1$). There are also 6 states of
     qqq'-type made from the $\overrightarrow{\rho }$ osc. ( and
     partial $qq$ spin $S = 0$), 6 states of $qqq'$-type made from the
     $\overrightarrow{\lambda }$ osc. ( and
     partial $qq$ spin $S = 1$). Finally, there are 4 states of
     $qq'q"$-type respectively made with the $\overrightarrow{\rho }$
and
$\overrightarrow{\lambda }$
     osc. and of $qq$ partial spin $S = 0$ and $1$.

     Then, with three flavours, one obtains a total of 70 states, with 8 states of $S = 3/2$
     and 19 states of $S = 1/2$ in perfect accordance with the usual
     SU(6) approach where the P-wave baryons ( L= 1 ) are classified in
     the irreducible 70 dimensional SU(6) representation,
     itself decomposing with respect to SU(3) flavour and SU(2) spin
     as:
     $$ 70 = ( 8, 3/2 ) + ( 8 + 10 + 1, 1/2 )$$
     In a similar manner, one can show that, for higher excitations,
     the number of excited states and their configurations are always
     what one should infer from the SU(6) approach.

\section{Flavour nonchanging forces}

\noindent Let us now focus on forces acting on the quarks (gluonic and
electromagnetic
in particular) that do not change the quark flavour.\\
 We can represent any baryon made of the three quarks
 $q_{1}$,$q_{2}$,$q_{3}$
as follows:
  \begin{eqnarray}\label{4}
        B(q_1,q_2,q_3) &=& (q_1 q_2)_s \otimes
        (q_3)_{1/2}
        \end{eqnarray}

\noindent  where $s$ is the spin of the
doublet
of the
quark pair $q_{1}q_{2}$. One obviously has $S=1$ or $S=0$.
\\
If we denote by $q_{i}^{\uparrow}$ the i quark with spin up, by
$q_{i}^{\downarrow}$ the corresponding
spin down state, then with the help of Clebsch-Gordan coefficients,
any $S=3/2$ state with $S_{z}=3/2$ writes :
 $$q_{1}^{\uparrow}q_{2}^{\uparrow}q_{3}^{\uparrow} $$
while it becomes for $S_{z}=1/2$ :
$$[ q_{1}^{\uparrow}q_{2}^{\uparrow}q_{3}^{\downarrow}
+q_{1}^{\uparrow}q_{2}^{\downarrow}q_{3}^{\uparrow}+
q_{1}^{\downarrow}q_{2}^{\uparrow}q_{3}^{\uparrow}
]/\sqrt{3}$$
and so on.
As we have seen there are 10 such flavour states, if we limit
ourselves to three flavours.\\
For the 8 flavour states with spin 1/2 and $S_{z}=+1/2$, one must
distinguish the case
  when the first two quarks are identical in flavour from the case
  when all flavours are different. In the first situation, the spin
of $q_{1}$ and $q_{2}$ must
couple to one ( we write below:
$q_{1}=q_{2}$ ) and Clebsch-Gordan coefficients give:
$$ \frac{1}{\sqrt{6}}\;[2 \cdot
q_{1}^{\uparrow}q_{1}^{\uparrow}q_{3}^{\downarrow}
-(q_{1}^{\uparrow}q_{1}^{\downarrow}+q_{1}^{\downarrow}q_{1}^{\uparrow})
q_{3}^{\uparrow}] $$. In the second configuration, the Pauli principle
gives no
restriction, and we have two states,
  the spin of the two first quarks being coupled to produce either a spin one
($\psi_1$)
   or zero ($\psi_2 $):\\
$$ \psi_1 = \frac{1}{\sqrt{6}}[2 \cdot q_{1}^{\uparrow}q_{2}^{\uparrow}q_{3}
^{\downarrow}
-(q_{1}^{\uparrow}q_{2}^{\downarrow}+q_{1}^{\downarrow}q_{2}^{\uparrow})
q_{3}^{\uparrow}]$$
$$\psi_2 = [(q_{1}^{\uparrow}q_{2}^{\downarrow}-
   q_{1}^{\downarrow}q_{2}^{\uparrow})
q_{3}^{\uparrow}]/\sqrt{2}$$
It clearly does not matter which ordering we have for the different
flavours,
but it will be shown in a moment that it is often convenient to order
them by
placing the lightest quark(s) in front of the heaviest.\\
We stress again that the $\it classification $ of states is into a
spin 3/2 decuplet and a spin 1/2 octet, $\it {no}$ $\it {matter}$
which three flavours we choose. With quarks d, c and b we have the
same structure as with the lightest states u, d and s.\\
 From this remark it should be clear for students that the
 classification of
states is one thing; the question  whether we have a flavour
symmetry group SU(3) is completely different.\\

Now, from the 8 states with total spin 1/2 that we have constructed, we
can span a Hilbert space and consider how interactions act on
states therein.\\
If there were no spin dependent interquark forces, it is evident that
the states made of the same quarks would have the same mass.
The interquark Hamiltonian would be diagonal in our Hilbert space.\\
This is definitely not so. In the beginning it seemed strange that the
$\Lambda$ and $\Sigma_{0}$ had different masses.
An early explanation, based on spin-spin interactions, was given by
Sakharov and Zeldowich in 1966 with the Hamiltonian :
\begin{equation}\label{5}
               H_S =   \sum_{i,j} c_{ij}
\overrightarrow{\sigma}_i
        \cdot \overrightarrow{\sigma}_j
\end{equation}
 all coefficients $ c_{ij}$ being equal.
In 1975 this spin-spin interaction
was shown to be a consequence of QCD by De Rujula, Georgi and Glashow:
\begin{equation}\label{1}
     H_{\mathrm{CM}} = - \sum_{i,j} C_{ij} \lala{i}{j} \SpSp{i}{j}
\end{equation}
As all quark pairs are in a $\overline{3}$ state of colour, the colour
part
  $\lala{i}{j} $ factorizes, giving a common factor - 8/3, so that:
\begin{equation}\label{1}
     H_{\mathrm{CM}} = (8/3) \sum_{i,j} C_{ij}  \SpSp{i}{j}
\end{equation}

\noindent Here the coefficients $C_{ij}$ are, among other things,
dependent
on the quark masses and properties of the spatial wave functions
of the quarks  in the system.\\
A natural scaling assumption for the coefficients
$ C_{ij} \propto \frac{1}{m_{i}m_{j}}$ comes from the analogy with the
hyperfine splitting in atoms that originates in the interaction between
(electro) magnetic moments.
In the physics of quarks  the analog interaction is between the
quarks $\it colourmagnetic$ moments.\\

Let us now imagine that we have integrated out all the spatial
variables
for the three quarks.
We can then write an effective Hamiltonian over the spin-space of the
quark:
\begin{equation}\label{5} \centering H =   \sum_{i} m_{i}  +
    (8/3) \sum_{i,j} C_{ij} \SpSp{i}{j}\end{equation}
Here effective masses $m_{i}$ incorporate the masses of the quarks as
well as their kinetic energy \footnote{If we added on
 $ H_S = -  \sum_{i,j} c_{ij}
\overrightarrow{\sigma}_i
        \cdot \overrightarrow{\sigma}_j $
\noindent we would get the most general Hamiltonian we can have
for
the system of three
  quarks when
the spatial variable are integrated out. So $H$ is more general than
$\it one$
  gluon exchange only.}.
If we $\it {assume}$ that each effective mass $m_{i}$ is (almost)
the same in different baryons, then we get mass formulae.
For this, we have only to determine the eigenvalues of
$H_{\mathrm{CM}}$.\\
The solution of the eigenvalue problem comes easily when one uses the
following
identity
for the Pauli matrices - as can be directly tested out by applying $
\overrightarrow{\sigma}_i
        \cdot \overrightarrow{\sigma}_j $ on the (symmetric) spin one
state and on the (antisymmetric) spin zero state:
\begin{equation}\label{1}
\centering  \overrightarrow{\sigma}_i
\cdot \overrightarrow{\sigma}_j   =-1+2 P_{i<->j} \end{equation}
 where $P_{i<->j}$ is the operator that
  permutes the spin states of the two particles i and j. One sees at once that
the Hamiltonian is almost diagonal in our Hilbert space.\\
For states with total spin $S = 3/2$ the eigenvalue
of $H_{\mathrm{CM}}$ is then:
\begin{equation}
  \centering \frac{8}{3} (\it C_{12}+ \it C_{13} +\it C_{23}  )
  \end{equation}
while for the spin S = 1/2 baryons, it reduces to :
\begin{equation}\label{13}
  \centering \frac{8}{3} (\it C_{12} -4 \it
C_{13})
\end{equation}
for all states with two identical quarks (they are chosen above as
being $q_1$ and $q_2$). The mass of corresponding S=1/2 states
reads then:
\begin{equation}\label{10}
  \centering 2m_1+m_3 +
\frac{8}{3} (\it C_{12}- 4 \it C_{13} )
\end{equation}
The only two states that are mixed are the spin 1/2 states where
all three flavours are different. Using the same spin coupling
scheme as before, where the  spin of the first two quarks is
coupled to one ($\psi_1$ ) or zero  ($\psi_2$ ), and where the
total spin is 1/2:
\begin{eqnarray}
\psi_1 &=& |(q_1 q_2)_1 \rangle \otimes |
(q_3)_{1/2} \rangle \nonumber \\
\psi_2 &=& |(q_1 q_2)_0\rangle \otimes |
(q_3)_{1/2} \rangle
\end{eqnarray}
one easily finds the colour-spin Hamiltonian over
these two states as:
\begin{equation}\label{9}
H_{\mathrm{CM}}= \frac{8}{3}\; \left[ \begin {array}{cc}{\it
C_{12}}- 2 ({\it C_{13}} +{\it
C_{23}}) & - \sqrt {3} \left( {\it C_{13}}-{\it C_{23}} \right)
\\\noalign{\medskip}
     -\sqrt {3} \left( {\it C_{13}}-{\it C_{23}} \right) & -3
      {\it C_{12}}
     \end {array}\right ]\end{equation}
\medskip
Now we see that if $\it C_{13}$ = $\it C_{23}$ this matrix is also
diagonal.
If $q_1$ is the $u$-quark and $q_2$ is the $d$-quark, we
would expect this to be approximately
true. The same remark holds for the effective masses $m_{1} \simeq
m_{2}$.
Then we note that in this approximation the largest eigenvalue leads
to a mass:
\begin{equation}\label{10}
  \centering 2m_1+m_3 +
\frac{8}{3} (\it C_{12}- 4 \it C_{13} )
\end{equation}
and that is the mass of both the $uuq_3$ and
$ddq_3$
systems given by expression (11). Therefore we have  $\it {three}$ states
in the
octet made from quarks u, d and
$q_3$
($q_3$ being s, c or b), with the same mass ( as they
would have if we  had an isospin 1 state and isospin invariance
in
nature ) .\\
The approximate isospin independence that sits in the QCD
Lagrangian - due
to the
  smallness of the (current) $u$ and $d$ mass compared to $ \Lambda_{QCD}
  $ , -
reappears
  in the masses of the three-quark bound states.\\
  If we make the baryons out of the lightest quarks $u$, $d$ and $s$, we
see
that
the states we
  called $\psi_{1} $ and $\psi_{2} $ are those commonly denoted
$\Sigma_{0}$ and $\Lambda $.\\
  The mixing between $\psi_{1} $ and $\psi_{2} $ - in this case called
$\Lambda $ -
$\Sigma_{0} $ mixing -  is induced by an
( isospin breaking )
  inequality $\it C_{13} \neq \it C_{23}$.\\
So this type of mixing is the general one for all $\psi_{1} $ and
$\psi_{2}
$ states, it is quite small
for $uds$, $udc$, $udb$, much bigger for states like $ucb$.\\
In the general case the eigenvectors for $H_{cm}$ are :
$$V_{+} = cos{\theta} \cdot \psi_{1}   + sin{\theta} \cdot
\psi_{2}$$
$$V_{-} = -sin{\theta} \cdot  \psi_{1} + cos{\theta} \cdot
\psi_{2}$$
with corresponding eigenvalues for $H_{cm}$:
$$\lambda_{\pm }= \frac{8}{3} \;[-({\it C_{12}}+{\it C_{13}}+{\it
C_{23})
\pm  2\,
\sqrt {{\it {C_{12}}^{2}-{\it C_{12}}\,{\it C_{13}}-{\it C_{12}}\,{\it
C_{23}}+ {\it C_{13}}}^{2}-{\it C_{13}}\,{\it C_{23}}+ {\it
C_{23}}^{2}}\;}] $$
and mixing angle $\theta $ given by
\begin{equation}\label{5}
\centering \tan {(2 \cdot \theta)} =
 \frac{- \sqrt {3} \cdot (\it C_{13}-\it C_{23})}
{2 \cdot \it C12-(\it C13+\it C23)}
\end{equation}
For light ($q=u$, $d$
and $s$) quarks, baryon masses are well reproduced with masses
\begin{equation}\label{ing:eq:masslb}
  m_q=360,\quad m_s=535\,\mathrm{MeV},
\end{equation}
and  strength factors 
\begin{equation}\label{ing:eq:parlb}
   C_{qq}=18.5,\quad C_{qs}=12.5,\quad C_{ss}=9.5\,\mathrm{MeV}.
\end{equation}

  If one accepts that effective masses could differ when heavy quarks
   are present, a possible choice for S-wave charmed
baryon states could  be
\begin{equation}\label{ing:eq:massch}
m_c=1550,\quad  m_q=450,\quad
m_s=590\,\mathrm{MeV},
\end{equation}
with associated  strength factors
\begin{eqnarray}\label{ing:eq:parch}
&C_{qq}=20,\quad
C_{qc}=5,\quad &C_{qs}=15,\quad\nonumber\\
&C_{ss}=10,\quad
C_{cs}=4,\quad&C_{cc}=4\ \ \mathrm{MeV}.
\end{eqnarray}

\noindent The states we have constructed are considerably simpler than
those one often \footnote {But not always !} finds in
  textbooks and lecture notes. No mention is here of "mixed
symmetry octet states" or "generalized Pauli principle". The
states are symmetrized only where the Pauli principle demands
  it, and can be used for all calculations where flavour is not
changed.\\
  If one wants to complicate calculations by symmetrizing in all
three flavours, one is of course free to do so \footnote {In this
case our notation is not optimal. The coefficients $C_{ij}$ should
then be labeled by the quark flavours and not by numbers.}.\\
  Mathematically there is $\it {indeed}$ a one to one correspondence
  between the states that we have used and the states  that
are symmetrized in
  all three particles. But to ask a student to compute the magnetic
moment of the nucleons
  by using nine terms in the state, instead of three, is not very kind.
  That calculation is very simple indeed, using the magnetic
  moment operator for a baryon:
 $$ \overrightarrow{\mu} = \sum_{i} Q_i \cdot \mu_i\cdot
\overrightarrow{\sigma}_i $$

where in the definition of
  $\mu_i $ we have taken out the electric charge factor $Q_i$   of the quark
i,
but not the
  expected  $\frac{1}{m_i \cdot c}$ dependence.All the $\mu_i$'s are
  therefore positive.\\
The expectation value of $ \overrightarrow{\mu}$ for a
$|B,S=1/2,S_{z}=1/2
\rangle $
state, composed of  two identical quarks $\alpha $   and a different
one $\beta $ gives the value of the magnetic moment of the baryon B:
\begin{center}
 $\mu_B $ =$\langle B,1/2,1/2|\mu_z| B,1/2,1/2 \rangle $ =
$\frac{4}{3}
\cdot Q_{\alpha} \cdot \mu _{\alpha}$ -
  $\frac{1}{3} \cdot Q_{\beta} \cdot \mu _{\beta}$ \\
  \end{center}
  With these conventions one gets for the proton and neutron:
  $ \mu_P =\frac{8}{9} \mu_u  +\frac{1}{9} \mu_d $,
  $ \mu_N =- \frac{2}{9} \mu_u  -\frac{4}{9} \mu_d $ and so on.\\
  In the case where all three flavours are different, $ q_1 =\alpha $,
$ q_2 =\beta $ and $ q_3 =\gamma $, one finds for the two states

$$\begin{array}{c}\label{ing:eq:magn}
\mu_{\psi_1}    = \langle \psi_1|\mu_z | \psi_1 \rangle = (2\cdot
Q_{\alpha}\cdot \mu_{\alpha} +2\cdot Q_{\beta }\cdot \mu_{\beta}
-Q_{\gamma }\cdot \mu_{\gamma})/3
\\[2mm]
\mu_{\psi_2} = \langle \psi_2|\mu_z| \psi_2 \rangle =Q_{\gamma
}\cdot \mu_{\gamma}\end{array} $$
\noindent We include the
off-diagonal matrix element for those that want to compute the
effect of $\psi_1 $ $ \psi_2 $ mixing:
$$ \langle \psi_2|\mu_z| \psi_1 \rangle
  = \langle \psi_1|\mu_z| \psi_2 \rangle = (-Q_{\alpha}\cdot
\mu_{\alpha}
  +Q_{\beta }\cdot \mu_{\beta} )/{\sqrt{3}} $$
  Neglecting small mixing, the magnetic moment of $\Lambda $ is given as
  $ \mu_{\Lambda} =- \frac{1}{3} \mu_s$. A tolerable estimation of all
  the magnetic moments of the lightest baryons can be obtained by using
  as input the observed nucleon and $\Lambda $ values, i.e.
  $\mu_u =\mu_d = \mu_P  , \mu_s = -3 \cdot \mu_{\Lambda}$.\\
  The student will remark that $\mu_s < \mu_u $ as expected, but also
  that not all is well with this kind of calculation. An obvious problem is
that
  experimentally $\mu_{\Lambda} - \mu_{\Xi^{-}}$ is positive, whereas
  this conventional type of calculation gives $\mu_{\Lambda} -
\mu_{\Xi^{-}}$=
  $ (\mu_s - \mu_d )/9$ which should be negative!\\
  We now turn to interactions that change flavour, and we shall see
that permutation symmetry
  reestablishes something equivalent to the "generalized Pauli
principle" without invoking any
  "principle", just the consistency of the Hilbert space.

\section{Flavour changing forces and the emergence of the generalized
Pauli principle.}

Suppose now that we want to teach flavour changing weak decays, as it
is the case in semi-leptonic hyperon decays.
Let us take the example of an s-quark  turning into a u-
quark as in the $\Lambda \rightarrow {P}{e}^{-}\bar{\nu}$ decay.\\
The state  $\Lambda $ is constituted by three quarks of different
flavours: u, d, and s, so the Pauli principle does not
bring any constraint in this case. This resonance  has a total spin S
= 1/2,
and its (ud) part is of spin S = 0 when we ignore $\Lambda \Sigma_0 $ mixing.
 We are therefore inclined to write it as:
\begin{center}
 $\Lambda
  = (u^{\uparrow}d^{\downarrow}-u^{\downarrow}d^{\uparrow})
  s^{\uparrow}/\sqrt{2}$
\end{center}
In the transformation $ {s}\rightarrow {u}$, the above expression
changes into:
\begin{center}
 $ (u^{\uparrow}d^{\downarrow}-u^{\downarrow}d^{\uparrow})
  u^{\uparrow}/\sqrt{2}$
\end{center}
and this state is not in the Hilbert space defined in the previous
section!
   It is not a state  allowed by the Pauli principle.\\
   The result is a state vector where the first term is symmetric under the interchange of the two u-quarks
-and that is fine- but the second term is not! As a consequence,
our Hilbert space is not appropriate to admit the action of
Hamiltonian
corresponding to the weak $|\Lambda \rangle $ decays.\\
Looking at the list of semi-leptonic decays of low-lying spin
1/2 baryons, one realizes that the same kind of pathology is
present as soon as the state  $|abc\rangle$ is transformed into
the state $|aac\rangle$, or inversely when the state $|aac\rangle$ is
transformed into a combination of $|abc\rangle$ states, with {a}, {b}
and {c}
figuring out the quark flavours.
But we see that this problem can easily be cured in the
above considered $|\Lambda \rangle $ decay by choosing for $|\Lambda
\rangle $
a wave function symmetric in u and s :
  \begin{center}
 $\Lambda = (u^{\uparrow}d^{\downarrow}s^{\uparrow}+
 s^{\uparrow}d^{\downarrow}u^{\uparrow}-u^{\downarrow}d^{\uparrow}s^{\uparrow}
 -s^{\uparrow}d^{\uparrow}u^{\downarrow} ) / 2$
\end{center}
and in the general case by
imposing the baryon wavefunctions to be completely symmetric
in all the three quarks whether they were identical or not. The
necessity of having a "complete" Hilbert space is what leads to the
"generalized" Pauli principle.\\
Actually, it can be shown that the Hilbert space description in terms of
completely
symmetrized states in the three quarks is equivalent to the
description in terms of
states where the symmetry is imposed only for quarks of the same
flavour. In other words, there is an isomorphism between the physics
described in terms of two
Hilbert spaces, the first one submitted to the usual Pauli principle
and the second
one submitted to what has been called the "generalized " Pauli
principle. The mathematical proof is sketched in Appendix B.\\
Let us point out that, instead of the commonly adopted
representation for the baryon wave-function ( we call it
"Generalized Pauli symmetric" in Appendix D ), one might prefer
the equivalent, simpler and more elegant expressions offered by
the Fock space formalism, as developed in Appendix C and
explicated in Appendix D for the octet of $S=1/2$
low-lying baryons.\\
Finally, combining the completely antisymmetric colour part
of the three quark wave functions with its completely symmetric
complementary part provides, of course, with a wave function which is
completely
antisymmetric in the three constituent fermions, and that is namely
what is
prescribed by the generalized Pauli principle.

\section{Conclusion}
We have shown that the classification of baryons, and reasonable
estimates of
their masses, their magnetic moments, as well as of the corresponding
form factors relative to semi-leptonic decays can be obtained without
any explicit reference to the flavour unitary group. All this can be
included as well in
fairly elementary quantum mechanics courses, as in elementary
particle courses.\\
As widely explicated, the "minimal" or "partial" permutational
symmetry imposed on wave-functions by the Pauli principle is well
adapted as long as flavour changing forces do not operate. When
they do, one is naturally led to the choice of wave functions
satisfying complete symmetry under permutation of the three
constituent quarks, that is, in other words,
to the application of the generalized Pauli principle. \\
At this point it is interesting
to make a connection between the
above discussion in Section 5 and the classification of baryons in
irreducible
representions of the SU(N) group, N being the number of different
flavours. Actually, one knows that each irreducible representation
$\mathcal{R}$ of
the SU(N) group can be represented by a Young tableau with n boxes
associated with an irreducible representation of the symmetric group
$\mathcal{S}_{n}$ also called permutation group $\mathcal{P}_{n}$
of n elements. That means that any element of $\mathcal{R}$ satisfies
a special $\mathcal{P}_{n}$ symmetry which is the same for all
elements in $\mathcal{R}$. Taking as an example SU(2), the
representations above denoted S=3/2 and S=1/2 resulting from the
tensorial product of three times the fundamental representation S=1/2
are respectively completely symmetric and "mixed symmetric"
under $\mathcal{P}_{3}$. Now, considering the group SU(3), its
irreducible eight dimensional  $\underline 8$ and ten dimensional
$\underline {10}$ representations
obtained by tensorial product of three  fundamental
representations $\underline 3$ are respectively completely and mixed
symmetric under
 $\mathcal{P}_{3}$. Complete symmetry under flavour and spin will
 therefore be obtained by combining the spin-SU(2) and flavour-SU(3)
 representations as follows: (S=1/2; $\underline 8$) and (S=3/2;
$\underline
 {10}$). And we can here recognize the decomposition of the
 completely symmetric $\underline{56}$ representation of the
 so-called flavour-spin SU(6) group commonly used to classify the
 S-wave baryons (in perfect accordance with results of Section 2).
 That also makes unitary groups so natural candidates for (broken)
 symmetry groups.

\section*{Acknowledgements :}
P.S. is indebted to P. Aurenche and E. Guadagnini for numerous and
valuable discussions. We both must specially thank R. Stora for his
precious help in some algebraic aspects developed in this note,
particularly in Appendices B and C, as well as  for a critical reading
of the manuscript.

\section*{Appendix A: Jacobi coordinates }
The spatial wave functions $\psi$ of our problem are the
harmonic-oscillator eigenfunctions
$\psi_{nLM}(\overrightarrow{\rho},\overrightarrow{\lambda}$), where
$\overrightarrow{\rho}$ and $\overrightarrow{\lambda}$
are the Jacobi coordinates which separate the Hamiltonian in
Eq.~(\ref{IKHam}) into two independent three-dimensional
oscillators.\\
Hereafter, we give expressions for Jacobi coordinates that makes the
separation
of the center of mass motion in the case of unequal masses, and even
when the
interaction potential is different for each pair of particles.\\
So, starting from the Hamiltonian:\begin{equation}
H_{\rm h.o.}=\sum_i (m_i + p_{i}^2 / 2m_i)
+ \sum_{i<j}  (k_{ij}/2) \cdot (\overrightarrow r_i -\overrightarrow
r_j)^2
\label{IKHam}
\end{equation}
with each coupling $k_{ij}= k_{ji}$ explicitly depending of the quark
pair $q_{i}q_{j}$,
one can rewrite the potential part as:
$$\sum_{i<j}  (k_{ij}/2) \cdot (\overrightarrow r_i -\overrightarrow
r_j)^2 = 1/2 \;\overline Y \mathcal{M} Y$$
with:
$$\overline Y= (\overrightarrow{y_{1}},\overrightarrow{y_{2}},
\overrightarrow{y_{3}})$$
and  \overrightarrow{y_{i}} defined as: $\overrightarrow{y_{i}}=
\sqrt{m_{i}} \overrightarrow{r_{i}}$, and the entries of the matrix
$\mathcal{M}$ satisfying:
  $$\mathcal{M}_{ii}=\frac{k_{ij}+k_{ik}}{m_{i}}\;\;\;i\neq{j}\neq{k}$$
 $$\mathcal{M}_{ij}=\mathcal{M}_{ji}= -\frac{k_{ij}}
 {(m_{i}m_{j})^\frac{1}{2}} \;\;\;\; i\neq{j}$$
 The diagonalization of $\mathcal{M}$ provides with one eigenvalue
  $\lambda_{0}=0$ associated to an eigenvector proportional to:
  \begin{equation}\overrightarrow{R_{{cm}}} = \frac {1}{M}
(m_{1}\overrightarrow r_1
+ m_{2} \overrightarrow r_2 + m_{3}\overrightarrow r_3 )
\end{equation}
where: $M= m_{1}+ m_{2}+ m_{3}$, and two non zero eigenvalues
$\lambda_{\pm}$ respectively associated to the eigenvectors:
\begin{equation}\overrightarrow{\rho } =
h_{+}(m_{1}^{1/2}\overrightarrow r_1
+ m_{2}^{1/2}f_{+}\overrightarrow r_2 +
m_{3}^{1/2}g_{+}\overrightarrow r_3)\end{equation}
and:
\begin{equation}\overrightarrow{\lambda } =
h_{-}(m_{1}^{1/2}\overrightarrow r_1
+ m_{2}^{1/2}f_{-}\overrightarrow r_2 +
m_{3}^{1/2}g_{-}\overrightarrow r_3)\end{equation}
with:
$$h_{\pm}=
\sqrt{\frac{\lambda_{\pm}}{(3K)^\frac{1}{2}(1+f_{\pm}^{2}+g_{\pm}^{2})}}$$
$$K = k_{12}k_{13} +  k_{21}k_{23}+ k_{31}k_{32}$$
and:
$$f_{\pm} = \sqrt{\frac{m_{2}}{m_{1}}} \; \frac{[k_{23}+k_{31} -
m_{3}\lambda_{\pm})] [k_{31}+k_{12}-m_{1}\lambda_{\pm}] -
k_{31}^{2}}{k_{23}k_{31} +
k_{12}[k_{23}+k_{31}-m_{3}\lambda_{\pm}]}$$
$$g_{\pm} = \frac{\sqrt{m_{3}}}{k_{23}} \; [ -
\frac{k_{12}}{\sqrt{m_{1}}} +
\frac{k_{23}+k_{12}-m_{2}\lambda_{\pm}}{\sqrt{m_{2}}}
f_{\pm}]$$

\medskip

Finally, the explicit expressions of $\lambda_{\pm}$  read:

$$\lambda_{\pm}= 1/2 \; [ \sum_{i;\; i\neq{j}\neq{k}\neq{i}}
\frac{k_{ij}+k_{ik}}{m_{i}}\pm \sqrt{ \left(\sum_{i;\;
i\neq{j}\neq{k}\neq{i}} \frac{k_{ij}+k_{ik}}{m_{i}}\right)^{2} - 4
K\sum_{j<k}\frac{1}{m_{j}m_{k}}} ]$$

\medskip

In terms of $\overrightarrow{R_{{cm}}},\overrightarrow{\rho
},\overrightarrow{\lambda }$, the total Hamiltonian acquires the
following simple form in which the c.m. motion explicitly separates:
\begin{equation}
H_{\rm h.o.}=\sum_i m_i + \frac{p_{R_{cm}}^2}{2M} +
\frac{p_{\rho}^2}{ 2m_{+}} + \frac{p_{\lambda}^2 }{ 2m_{-}} +
\frac{\sqrt{3K}}{2} (\overrightarrow{\rho}^2 +
\overrightarrow{\lambda}^2)
\end{equation}
where: $m_{\pm}= \frac{\sqrt{3K}}{\lambda_{\pm}}$

\medskip

It might be useful to consider the special case  with only one
coupling constant, that is: $ k_{ij}= k$ for i,j= 1,2,3. Then, K
becomes  $= 3k^{2}$ and, in the Hamiltonian, the potential part
reduces to: $\frac{3}{2} k (\overrightarrow{\rho}^2 +
\overrightarrow{\lambda}^2)$. In the kinetic part, one gets:
$m_{\pm}=\frac{3k}{\lambda_{\pm}}$ and :
$$\lambda_{\pm}= k \;[\; \sum_{i} \frac{1}{m_{i}}\;\pm
\;\sqrt{\sum_{i}\frac{1}{m_{i}^{2}} -
\sum_{i<j}\frac{1}{m_{i}m_{j}}}\;]$$
while $f_{\pm}$ and $g_{\pm}$ simplify as :
$$f_{\pm}= \sqrt{\frac{m_{2}}{m_{1}}}
\;\frac{4[1-m_{3}(\lambda_{\pm}/2k)][1-m_{1}(\lambda_{\pm}/2k)]-1}
{3-2m_{3}(\lambda_{\pm}/2k)}$$
$$g_{\pm} = \sqrt{m_{3}}\;[ \;-\frac{1}{\sqrt{m_{1}}}
+2\frac{1-m_{2}(\lambda_{\pm}/2k)}{\sqrt{m_{2}}}
f_{\pm}]$$

\medskip

Another relevant case is the one with two identical quarks, that is
$m_{1}=m_{2}=m$ and $m_{3}=m'$. Then it is reasonable to have :
$k_{31}=k_{23}= k'$ and $k_{12}=k$, and the $ \overrightarrow{\rho
}$ and $\overrightarrow{\lambda }$ Jacobi vectors simply become:
 \begin{equation}\overrightarrow{\rho }= \frac{1}{\sqrt{2}}\;
 [\frac{2k+k'}{3k'}]^\frac{1}{4}\; (\overrightarrow r_{1} -
\overrightarrow r_{2})\end{equation}
\begin{equation}\overrightarrow{\lambda }= \frac{1}{\sqrt{2}}\;
 [\frac{k'}{3(2k+k')}]^\frac{1}{4}\; (\overrightarrow r_{1} +
\overrightarrow r_{2} -2\overrightarrow r_{3})\end{equation}

.
\medskip

Finally, we give the limit case with all three masses equal
$m_{1}=m_{2}=m_{3}=m$ and $k_{12}=k_{23}=k_{31}=k$ :
\begin{equation}\overrightarrow{\rho }= \frac{1}{\sqrt{2}}\;
 (\overrightarrow r_{1} -
\overrightarrow r_{2})\end{equation}
\begin{equation}\overrightarrow{\lambda }= \frac{1}{\sqrt{6}}\;
 (\overrightarrow r_{1} +
\overrightarrow r_{2} -2\overrightarrow r_{3})\end{equation}
\vspace{8mm}

\section*{Appendix B:  Generalized Pauli principle versus Pauli
principle: a mathematical proof of their equivalence }
This section is not directly dedicated to students. Of course, one
can try and convince oneself on some examples that there is a
one-to-one
correspondence between the "mathematical" states and observables
submitted to the
Generalized
Pauli principle (or GPP) and those simply satisfying the Pauli
principle (or PP). The proof that we propose hereafter has two
advantages. First, it is general, and so adaptable to  situations
other
than that of regular baryons (for example multiquark states).
Secondly, it introduces the Fock space formalism in a rather natural
way.  All that follows is taken from \cite{Bourbaki}.
We start with some definitions :\\
Let $k$ be a commutative ring, $E$ a $k$-module, and $N$ the set of
integers.\\

$\bf{Def.1}$: Symmetric tensor algebra:

Denoting $T^n (E)$ the set of elements: ${z}={x_{1}}\otimes{
x_{2}}\otimes\ldots
\otimes{x_{n}}$
with $x_{1},\ldots,x_{n }\in M$, the action of $\mathcal{S}_{n}$
on $T^n (E)$ reads:
\begin{equation} \sigma({x_{1}}\otimes{ x_{2}}\otimes\ldots
\otimes{x_{n}})= x_{\sigma^{-1}(1)} \otimes{
x_{\sigma^{-1}}(2)}\otimes\ldots \otimes x_{\sigma^{-1}(n)}
\end{equation}

The elements ${z}$ such that:
$$\sigma\cdot{z}=z$$
are called symmetric tensors of order n. They form a
sub-$\mathit{k}$-module of  $T^n (E)$, denoted: $TS^n (E)$. One sets:
$$TS(E)=\bigoplus_{n=0}^{\infty}{TS^n(E)}$$
on which one can  define a symmetric product (we do not give here the
rule in order not to overload the text).\\

$\bf{Def.2}$: Gamma algebra:

We call Gamma algebra of E and denote by $\Gamma(E)$ the associative,
unifier, commutative, algebra defined by the set of generators
${N}\times{E}$ and relations:
\begin{equation} (0,x)= 1 \end{equation}
\begin{equation}(p,\lambda x)= \lambda^p (p,x) \end{equation}
\begin{equation}(p,x+y)= \sum_{q=0}^p (p,x) (p-q,y) \end{equation}

$\bf{Def. 3}$: Exponential type sequence:

We call exponential type sequence of E elements
 a sequence $a= (a_{p})_{p\in{N}}$ such that
 \begin{equation}a_{0}=1 \end{equation}
  \begin{equation}a_{p}a_{q}=\frac{(p+q)!}{p!q!} \;a_{p+q}
  \end{equation}\\
 Now, we come to the results we need:

  $\bf{ Prop.1}$:   $\lbrace{(p,x)}\rbrace_{p\in{N}}$ with $x\in{E}$
is an
 exponential type sequence.
 \medskip

  An interesting example of exponential type sequence is given by
 $f(x)$  with
 \begin{equation}f(x)_{p}= \frac{1}{p!} \;(x)^{p}\end{equation}

 $\bf{Prop.2}$: There exists one and only one isomorphism g between
$\Gamma(E)$
 and $TS(E)$:

 \begin{equation} \Gamma(E)\cong{TS(E)}\end{equation} such that:
 \begin{equation}({g}\circ{\gamma_{p}})(x)={x}\otimes{ x}\otimes\ldots
\otimes{x}\end{equation}
with p $x$-factors, and where we have denoted by $\gamma_{p}$, with
$p\in{N}$,
the application from  $E$
into $\Gamma(E)$ product of the injection
$x\to (p,x)$ and of the canonical homomorphism of the free
commutative
algebra ${N}\times{E}$ to $\Gamma(E)$.\\

$\bf{Prop.3}$: Let $E$ and $F$ be two ${k}$ modules.  There exists
one and
only one isomorphism $\phi$ from $\Gamma({E}\times{F})$ into
$\Gamma({E})\otimes_{k}\Gamma({F})$:

\begin{equation}\Gamma({E}\times{F})\cong{\Gamma({E})\otimes_{k}\Gamma({F})}
\end{equation}
such that:
 \begin{equation}(\phi\circ{\gamma_{p}})(x,y)=\sum_{q=0}^p
\gamma_{q}(x)\otimes\gamma_{p-q}(y) \end{equation}
\medskip

 The above properties can be reformulated by considering the
 case of polynomials. Indeed, one directly remarks that $\Gamma(E)$
can
 be seen as the set of polynomials on the  dual $E^{*}$ of $E$, that
is also:
\begin{equation}\Gamma(E)\cong{Polyn( E^{*})}\end{equation}
Now, considering E  finite dimensional, that is made of elements
$\overrightarrow{x}=  \sum_{i}x^{i}e_{i}\in{E}$ where
 $(e_{i})_{i\in{I}}$ is a basis of $E$,
  with $x_{i}\in{k}$ for each
$i\in{I}$, and $I$ being finite,
 then Prop.2 insures that:
\begin{equation}
\Gamma(E)\cong{Polyn(E^{*})}\cong{(Polyn(E))^{*}}\cong{TS(E)}
\end{equation}
  and from Prop.3 we get:
 \begin{equation} {Polyn(\overrightarrow{x},\overrightarrow{y})} =
{Polyn(\overrightarrow{x})}\otimes{Polyn(\overrightarrow{y})}\end{equation}
  with $\overrightarrow{x}\in{E}$ and $\overrightarrow{y}\in{F}$,
 keeping in mind the relations (31),(34) and (38).\\

Coming back to our flavoured quarks, let us associate to each
flavour f = u, d,\ldots  the $S=1/2$ spin representation space
$E_{f}$ generated by $f^{\uparrow},f^{\downarrow}$. One can write:
 \begin{equation}\Gamma(E_{f}) \cong {
     Polyn(f^{\uparrow},f^{\downarrow})}
 \end{equation}
and also :
 \begin{equation}\Gamma({E_{f}«}\times{E_{f'}«})
     \cong{\Gamma({E_{f}«})\otimes_{R}\Gamma({E_{f'}})}
\end{equation}
 our states being defined on the ring of real numbers R.

\section*{Appendix C: Fock space wave-functions}
Actually, the above developed framework is nothing else than Fock space.
Let us associate to each just above defined couple
$f^{\uparrow},f^{\downarrow}$
in $E_{f}$ the - commuting or "bosonic" - creator operators
$a^{\dag}_{f^{\uparrow}}$, $a^{\dag}_{f^{\downarrow}}$,
 acting on the $\vert{0}\rangle$ vacuum. Again, we can write:
 \begin{equation}\Gamma(E_{f}) \cong{ Polyn(a^{\dag}_{f^{\uparrow}},
a^{\dag}_{f^{\downarrow}}) }\end{equation}
and also:
\begin{equation} Polyn(a^{\dag}_{f^{\uparrow}},
a^{\dag}_{f^{\downarrow}})\otimes {Polyn(a^{\dag}_{f'^{\uparrow}},
a^{\dag}_{f'^{\downarrow}})}\cong{ Polyn(a^{\dag}_{f^{\uparrow}},
a^{\dag}_{f^{\downarrow}},a^{\dag}_{f'^{\uparrow}},
a^{\dag}_{f'^{\downarrow}})} \end{equation} Due to the invariance
under permutations of a monomial in $a^{\dag}_{f}$ operators, one
can get, for baryon states, expressions as simple as the ones used
in Section (4) and denoted in Appendix D below "simply Pauli
symmetric" - so much simpler than those called  "generalized Pauli
symmetric" in the same appendix - but unambiguously defined, as
these last ones, for computations involving flavour changes. More
precisely, it is rather immediate to translate a wave-function
written in our "Pauli symmetric basis" into the corresponding one
in the Fock basis. One has just to take care of the respective
normalization of states: as an example, the spin $S=1, S_{z}=0$
state $(u^{\uparrow}u^{\downarrow}+
u^{\downarrow}u^{\uparrow})/\sqrt{2}$ will correspond to
${a}_{u^{\uparrow}}^{\dag}{a}_{u^{\downarrow}}^{\dag}$, while
$u^{\uparrow}u^{\uparrow}$ will be associated to
${a}_{u^{\uparrow}}^{\dag}{a}_{u^{\uparrow}}^{\dag}/\sqrt{2}$,
using the usual normalized power product
$(a_{f}^{\dag})^{n}/\sqrt{n!}$. An explicit comparison of the
wave-functions for $S=1/2$ low-lying
baryons in the different bases can be performed with Appendix D.\\
\medskip
Finally, associating, as usual, to each creator $a_{f}^{\dag}$ its
annihilator counter part
$a_{f}$ such  that   $a_{f}\vert{0}\rangle = 0$ and
$(a_{f}^{\dag})^{\dag} = a$,
and satisfying the well-known
 Heisenberg algebra commutation relations:
\begin{equation}[a_{i},a_{j}^{\dag}] = \delta_{ij}\end{equation}
\begin{equation}[a_{i},a_{j}] = [a_{i}^{\dag},a_{j}^{\dag}] =
    0\end{equation}
    with i,j representing any $f^{\uparrow}$ or$f^{\downarrow}$,
  simple expressions can be given for the self-adjoint parts of the
 Hamiltonian representing the  flavour changing $f$ into $f'$:
 \begin{equation}H_{f\to{f'}}^{\pm} =
    ( a^{\dag}_{f'^{\uparrow}}{a}_{f^{\uparrow}}\pm
 a^{\dag}_{f'^{\downarrow}}{a}_{f^{\downarrow}} +
 h.c.)
\end{equation}
involved in the computations of the vector $f_{1}$ and axial $g_{1}$ form
 factors of baryon semi-leptonic decays. \\
 It might look surprising to the reader that we use a bosonic Fock
 space although we are dealing with fermionic quarks. We should not
 forget that antisymmetry is carried  by the colour part. In other
 words, let us say that in the fermionic space of coloured quarks,
 there
exists a
 subspace constituted by three quark states which are colour singlets
  - our baryons - this subspace being isomorphic, as a Hilbert
 space, to the three particles subspace of a bosonic Fock space.

    ,\section*{Appendix D: Wave-functions of spin 1/2 S-wave baryons }
 Different expressions for the wave-functions of
 baryons have been used following the kind of computations
 we had to consider. As long as classification and flavour non
 changing quantities are involved, wave-functions satisfying the
symmetry
 imposed by the Pauli principle are perfectly adapted: we will call
 them "partial symmetric" or "simple Pauli symmetric" wave-functions.
 As soon as flavour changing forces are involved, complete
 permutational invariance of the three quarks is required : we will
 denoted them "generalized Pauli symmetric" wave-functions. In this
 last case, the use of the Fock space formalism, as seen in Appendix
 C, looks rather appealing: their natural denomination will be "Fock"
 wave-functions. Let us once more mention that the colour part, and
 so the antisymmetry nature of the baryon, has been "factored out",
 that is why one is concerned below only with permutational symmetry
 aspects, and in particular with bosonic Fock operators.\\

 $\textbf{ Simple  Pauli symmetric  wave-functions}$

  \begin{center}
 ${N}= [2d^{\uparrow}d^{\uparrow}u^{\downarrow}-(d^{\uparrow}d^{\downarrow}+
 d^{\downarrow}d^{\uparrow})
  u^{\uparrow}]/\sqrt{6}$
\end{center}

 \begin{center}
 ${P}= [2u^{\uparrow}u^{\uparrow}d^{\downarrow}-(u^{\uparrow}u^{\downarrow}+
 u^{\downarrow}u^{\uparrow})
  d^{\uparrow}]/\sqrt{6}$
\end{center}

 \begin{center}
 $ \Lambda = [(u^{\uparrow}d^{\downarrow}-u^{\downarrow}d^{\uparrow})
  s^{\uparrow}]/\sqrt{2}$
\end{center}

 \begin{center}
 ${\Sigma^{O}}=
[2u^{\uparrow}d^{\uparrow}s^{\downarrow}-(u^{\uparrow}d^{\downarrow}+
 u^{\downarrow}d^{\uparrow})
  s^{\uparrow}]/\sqrt{6}$
\end{center}

 \begin{center}
 $\Sigma^{+}=
[2u^{\uparrow}d^{\uparrow}s^{\downarrow}-(u^{\uparrow}u^{\downarrow}+
 u^{\downarrow}u^{\uparrow})
  s^{\uparrow}]/\sqrt{6}$
\end{center}

 \begin{center}
 $\Sigma^{+}=
[2d^{\uparrow}d^{\uparrow}s^{\downarrow}-(d^{\uparrow}d^{\downarrow}+
 d^{\downarrow}d^{\uparrow})
  s^{\uparrow}]/\sqrt{6}$
\end{center}

\begin{center}
 $\Xi^{-}=
[2s^{\uparrow}s^{\uparrow}d^{\downarrow}-(s^{\uparrow}s^{\downarrow}+
 s^{\downarrow}s^{\uparrow})
  d^{\uparrow}]/\sqrt{6}$
\end{center}

\begin{center}
 $\Xi^{0}=
[2s^{\uparrow}s^{\uparrow}u^{\downarrow}-(s^{\uparrow}s^{\downarrow}+
 s^{\downarrow}s^{\uparrow})
  u^{\uparrow}]/\sqrt{6}$
\end{center}

$\textbf{ Generalized  Pauli symmetric  wave-functions}$

  \begin{center}
 ${N}= [2d^{\uparrow}d^{\uparrow}u^{\downarrow}-(d^{\uparrow}d^{\downarrow}+
 d^{\downarrow}d^{\uparrow})
  u^{\uparrow}+ permutations ]/\sqrt{18}$
\end{center}

 \begin{center}
 ${P}= [2u^{\uparrow}u^{\uparrow}d^{\downarrow}-(u^{\uparrow}u^{\downarrow}+
 u^{\downarrow}u^{\uparrow})
  d^{\uparrow}+ permutations]/\sqrt{18}$
\end{center}

 \begin{center}
 $ \Lambda = [(u^{\uparrow}d^{\downarrow}-u^{\downarrow}d^{\uparrow})
  s^{\uparrow} + permutations]/\sqrt{12}$
\end{center}

 \begin{center}
 $\Sigma^{O}=
[2u^{\uparrow}d^{\uparrow}s^{\downarrow}-(u^{\uparrow}d^{\downarrow}+
 u^{\downarrow}d^{\uparrow})
  s^{\uparrow} + permutations]/ 6$
\end{center}

 \begin{center}
 $\Sigma^{+}=
[2u^{\uparrow}d^{\uparrow}s^{\downarrow}-(u^{\uparrow}u^{\downarrow}+
 u^{\downarrow}u^{\uparrow})
  s^{\uparrow} + permutations]/\sqrt{18}$
\end{center}

 \begin{center}
 $\Sigma^{-}=
[2d^{\uparrow}d^{\uparrow}s^{\downarrow}-(d^{\uparrow}d^{\downarrow}+
 d^{\downarrow}d^{\uparrow})
  s^{\uparrow} + permutations]/\sqrt{18}$
\end{center}

\begin{center}
 $\Xi^{-}=
[2s^{\uparrow}s^{\uparrow}d^{\downarrow}-(s^{\uparrow}s^{\downarrow}+
 s^{\downarrow}s^{\uparrow})
  d^{\uparrow} + permutations]/\sqrt{18}$
\end{center}

\begin{center}
 $\Xi^{0}=
[2s^{\uparrow}s^{\uparrow}u^{\downarrow}-(s^{\uparrow}s^{\downarrow}+
 s^{\downarrow}s^{\uparrow})
  u^{\uparrow} + permutations]/\sqrt{18}$
\end{center}

$\textbf{ Fock wave-functions}$

\begin{center}
 ${N}=
\frac{1}{\sqrt{3}}\;[{a}_{d^{\uparrow}}^{\dag}{a}_{d^{\uparrow}}^{\dag}
{a}_{u^{\downarrow}}^{\dag}-{a}_{d^{\uparrow}}^{\dag}{a}_{d^{\downarrow}}^{\dag}
 {a}_{u^{\uparrow}}^{\dag}] \vert{0}\rangle$
 \end{center}

\begin{center}
 ${P}=
\frac{1}{\sqrt{3}}\;[{a}_{u^{\uparrow}}^{\dag}{a}_{u^{\uparrow}}^{\dag}
{a}_{d^{\downarrow}}^{\dag}-{a}_{u^{\uparrow}}^{\dag}{a}_{u^{\downarrow}}^{\dag}
 {a}_{d^{\uparrow}}^{\dag}] \vert{0}\rangle$
 \end{center}

\begin{center}
 ${\Lambda}=
\frac{1}{\sqrt{2}}\;[{a}_{u^{\uparrow}}^{\dag}{a}_{d^{\downarrow}}^{\dag}
{a}_{s^{\uparrow}}^{\dag}-{a}_{u^{\downarrow}}^{\dag}{a}_{d^{\uparrow}}^{\dag}
 {a}_{s^{\uparrow}}^{\dag}] \vert{0}\rangle$
 \end{center}

 \begin{center}
 ${\Sigma^{0}}= \frac{1}{\sqrt{6}}\;[
2{a}_{u^{\uparrow}}^{\dag}{a}_{d^{\uparrow}}^{\dag}
{a}_{s^{\downarrow}}^{\dag}-{a}_{u^{\uparrow}}^{\dag}{a}_{d^{\downarrow}}^{\dag}
 {a}_{s^{\uparrow}}^{\dag}
-{a}_{u^{\downarrow}}^{\dag}{a}_{d^{\uparrow}}^{\dag}
 {a}_{s^{\uparrow}}^{\dag}] \vert{0}\rangle$
 \end{center}

\begin{center}
 ${\Sigma^{+}}=
\frac{1}{\sqrt{3}}\;[{a}_{u^{\uparrow}}^{\dag}{a}_{u^{\uparrow}}^{\dag}
{a}_{s^{\downarrow}}^{\dag}-{a}_{u^{\uparrow}}^{\dag}{a}_{u^{\downarrow}}^{\dag}
 {a}_{s^{\uparrow}}^{\dag}] \vert{0}\rangle$
 \end{center}

\begin{center}
 ${\Sigma^{-}}=
\frac{1}{\sqrt{3}}\;[{a}_{d^{\uparrow}}^{\dag}{a}_{d^{\uparrow}}^{\dag}
{a}_{s^{\downarrow}}^{\dag}-{a}_{d^{\uparrow}}^{\dag}{a}_{d^{\downarrow}}^{\dag}
 {a}_{s^{\uparrow}}^{\dag}] \vert{0}\rangle$
 \end{center}

\begin{center}
 ${\Xi^{-}}=
\frac{1}{\sqrt{3}}\;[{a}_{s^{\uparrow}}^{\dag}{a}_{s^{\uparrow}}^{\dag}
{a}_{d^{\downarrow}}^{\dag}-{a}_{s^{\uparrow}}^{\dag}{a}_{s^{\downarrow}}^{\dag}
 {a}_{d^{\uparrow}}^{\dag}] \vert{0}\rangle$
 \end{center}

\begin{center}
 ${\Xi^{0}}=
\frac{1}{\sqrt{3}}\;[{a}_{s^{\uparrow}}^{\dag}{a}_{s^{\uparrow}}^{\dag}
{a}_{u^{\downarrow}}^{\dag}-{a}_{s^{\uparrow}}^{\dag}{a}_{s^{\downarrow}}^{\dag}
 {a}_{u^{\uparrow}}^{\dag}] \vert{0}\rangle$
 \end{center}

\end{document}